\documentclass[journal=nalefd,manuscript=letter,layout=traditional,citeautoscript]{achemso}

\setkeys{acs}{etalmode=truncate}
\setkeys{acs}{maxauthors=0}
\setkeys{acs}{articletitle=true}

\usepackage{amssymb}

\usepackage{epsfig}
\usepackage{graphicx}
\usepackage{amsmath}
\usepackage{bm}
\usepackage{color}
\usepackage{siunitx}
\usepackage[normalem]{ulem}

\usepackage[breaklinks]{hyperref}
\usepackage[all]{hypcap}
\hypersetup{
    plainpages=false,
    bookmarks=false,         
    unicode=false,          
    pdfborder={0 0 0}
    pdftoolbar=true,        
    pdfmenubar=true,        
    pdffitwindow=false,     
    pdfstartview={FitH},    
    pdftitle={BulkCrystalMagnetoOptics},    
    pdfauthor={Michal Baranowski},     
    pdfproducer={LNCMI-CNRS-UGA-UPS-INSA}, 
    pdfkeywords={High} {Magnetic} {Fields}, 
    pdfnewwindow=true,      
    linktoc=section,
    colorlinks=true,       
    linkcolor=blue,          
    citecolor=red,        
    filecolor=magenta,      
    urlcolor=blue           
}

\title{Moir\'e Intralayer Excitons in a MoSe$_2$/MoS$_2$ Heterostructure}

\author{Nan Zhang}
\altaffiliation{These authors contributed equally to the work}
\affiliation{Laboratoire National des Champs Magn\'etiques Intenses, UPR 3228, CNRS-UGA-UPS-INSA, Grenoble and Toulouse, France}

\author{Alessandro Surrente}
\altaffiliation{These authors contributed equally to the work}
\affiliation{Laboratoire National des Champs Magn\'etiques Intenses, UPR 3228, CNRS-UGA-UPS-INSA, Grenoble and Toulouse, France}

\author{Micha{\l} Baranowski}
\affiliation{Laboratoire National des Champs Magn\'etiques Intenses, UPR 3228, CNRS-UGA-UPS-INSA, Grenoble and Toulouse, France}
\altaffiliation{Department of Experimental Physics, Faculty of Fundamental Problems of Technology, Wroclaw University of Science and
Technology, 50-370 Wroclaw, Poland}

\author{Duncan K.\ Maude}
\affiliation{Laboratoire National des Champs Magn\'etiques Intenses, UPR 3228, CNRS-UGA-UPS-INSA, Grenoble and Toulouse, France}

\author{Patricia Gant}
\affiliation{Materials Science Factory, Instituto de Ciencia de Materiales de Madrid (ICMM), Consejo Superior de Investigaciones Científicas (CSIC), Sor Juana Inés de la Cruz 3, 28049 Madrid, Spain}

\author{Andres Castellanos-Gomez}
\affiliation{Materials Science Factory, Instituto de Ciencia de Materiales de Madrid (ICMM), Consejo Superior de Investigaciones Científicas (CSIC), Sor Juana Inés de la Cruz 3, 28049 Madrid, Spain}

\author{Paulina Plochocka}\email{paulina.plochocka@lncmi.cnrs.fr}
\affiliation{Laboratoire National des Champs Magn\'etiques Intenses, UPR 3228, CNRS-UGA-UPS-INSA, Grenoble and Toulouse, France}


\begin{document}

\begin{abstract}

Spatially periodic structures with a long range period, referred to as moir{\'e} pattern, can be obtained in van der Waals bilayers in the presence of a small stacking angle or of lattice mismatch between the monolayers. Theoretical predictions suggest that the resulting spatially periodic variation of the band structure modifies the
optical properties of both intra and interlayer excitons of transition metal dichalcogenides heterostructures. Here, we report on the impact of the moir{\'e} pattern formed in a
MoSe$_2$/MoS$_2$ heterobilayer encapsulated in hexagonal boron nitride. The periodic in-plane potential results in a splitting of
the MoSe$_2$ exciton and trion in both emission and absorption spectra. The observed energy difference between the
split peaks is fully consistent with theoretical predictions.

\end{abstract}

\maketitle

The vertical stacking of atomically thin planes of layered solids provides a rich playground to expand the properties of the
constituting layers, which gives rise to new and attractive features \cite{geim2013van,novoselov20162d}. The possibility to combine a plethora
of different layered materials allows to efficiently tailor the properties of van der Waals heterostructures. In particular, this approach has been successfully employed for transition metal
dichalcogenides (TMDs) \cite{dong2017progress,zhou20182d}. Sandwiching TMD monolayers between hexagonal boron nitride (hBN)
improves significantly their optical and electrical properties \cite{cadiz2017excitonic, fallahazad2016shubnikov}, paving the way
to access their rich excitonic and transport physics \cite{cadiz2017excitonic,fallahazad2016shubnikov, robert2018optical,
manca2017enabling, stier2018magnetooptics, chen2018superior, xu2017odd, movva2017density, bandurin2017high, wu2016even}. Bringing
TMD monolayers in close contact with graphene makes it possible to tune controllably the band gap of the TMD, owing to the
locally different dielectric environment \cite{raja2017coulomb}. Stacking different semiconducting TMDs also allows to overcome
the limitations of isolated TMD monolayers in valleytronic applications \cite{xiao2012coupled,xu2014spin, wang2012electronics,
mak2012control, zeng2012valley}, such as very short exciton and valley polarization lifetimes \cite{Lagarde2014, Wang2014,
Robert2016, Wang_urbaszek2015,Zhu2014}.  TMD heterostructures exhibit type II band alignment
\cite{kang2013band,dong2017progress,zhou20182d}, which leads to the formation of interlayer excitons with radiative and valley
lifetimes up to five orders of magnitude longer than for intralayer exitons \cite{Rivera2015,Rivera2016, nagler2017interlayer,
miller2017long, baranowski2017probing, surrente2018intervalley}. The helicity of the long lived interlayer exciton emission can
be further controlled by the polarization of the excitation laser \cite{Rivera2016,baranowski2017probing,surrente2018intervalley,
hanbicki2018double, hsu2018negative}, which makes van der Waals heterostrucures attractive for valleytronic applications.

Due to the weak van der Waals interactions in heterostructures, the lattice constant of each monolayer does not conform to that
of the underlying substrate. If monolayers with different lattice parameters or with a non-zero (but small) stacking angle are overlaid, a
moir\'e pattern is formed \cite{park2008anisotropic,xue2011scanning,yankowitz2012emergence,ponomarenko2013cloning,yu2017moire,
zhang2017interlayer}. The resulting in-plane superlattice potential has a tremendous impact on the electronic properties of the
van der Waals heterostructures, opening up new directions for material engineering, which relies on the relative orientation of
the constituting layers. The physics related to the moir\'e pattern has been studied in hBN/graphene heterostructures, where the
induced periodic potential leads to the formation of new Dirac cones, opening of a band gap, and the appearance of Hofstadter
butterfly states \cite{hunt2013massive, wang2016gaps,yankowitz2012emergence, ponomarenko2013cloning, dean2013hofstadter}. The
moir\'e pattern formed in TMD heterostructures is also expected to have a large influence on their properties \cite{yu2017moire,
zhang2017interlayer, wu2018theory, Wu2017}. According to theoretical predications, the moir\'e pattern should result in a
periodically modulated potential with minima for both intra \cite{Wu2017,yu2017moire} and interlayer \cite{yu2017moire,wu2018theory}
excitons, thus forming an array of quantum dots. Moreover, spatially varying selection rules for interlayer excitons can result in an emission with
helicity opposite with respect to that of the excitation laser \cite{yu2017moire,wu2018theory}. Finally, the variation of the
confinement potential for different atomic registries should lead to an energy splitting of both the intra \cite{Wu2017} and
interlayer transitions \cite{yu2017moire,wu2018theory}. The formation of a moir\'e pattern and the related periodic potential
fluctuations have been confirmed recently using scanning tunneling microscopy \cite{zhang2017interlayer,pan2018quantum}. However,
direct experimental evidence of the influence of the moir\'e pattern on the optical properties of TMD heterostructures has
remained elusive so far. The only indirect indication is related to the observation of counter polarized emission of the
interlayer exciton \cite{baranowski2017probing,hanbicki2018double,surrente2018intervalley,ciarrocchi2018control}.

In this work, we show how the observed splitting of the intralayer exciton and trion in a monolayer MoSe$_2$ assembled in a
heterostructure with MoS$_2$ and encapsulated in hBN is a direct consequence of the moir\'e pattern formed between MoSe$_2$ and
MoS$_2$. The high quality of our heterostructure (typical full width at half maximum of the exciton and trion photoluminescence, PL, peaks $\sim
\SI{5}{\milli\eV}$) allows us to reveal the splitting of the trion and exciton lines both in PL and
reflectivity spectra. The structure of the intralayer exciton transitions can be observed consistently over the whole area where the
two materials overlap, while we observed no splitting out of the heterostructure area. The energy splitting of the peaks and their
temperature dependence are in agreement with the expected influence of the moir\'e potential on the intralayer exciton species.
Our results provide a clear optical fingerprint of the effect of the moir\'e potential on intralayer excitons of a monolayer TMD.

\begin{figure*}[!ht]
\centering
\includegraphics[width=1\linewidth]{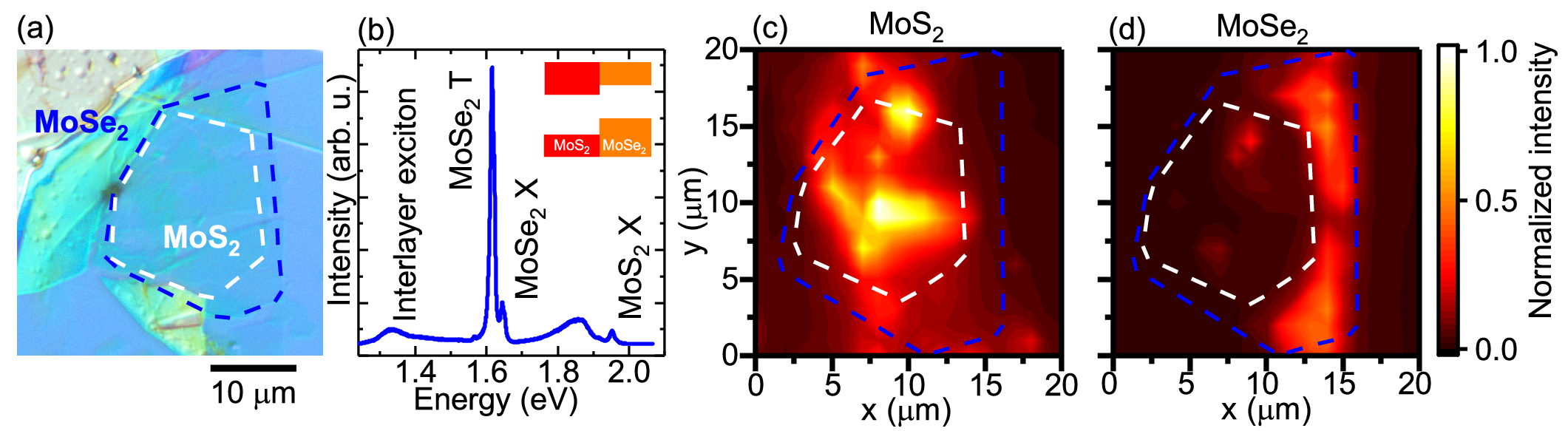}
\caption{(a) Optical microscope image of the MoS$_2$/MoSe$_2$ heterostructure encapsulated in hBN. The blue and white dashed
lines indicate the contours of the MoSe$_2$ and MoS$_2$ flakes. (b) Broad range PL spectrum of the heterostructure. Inset: schematic band alignment of a MoSe$_2$/MoS$_2$ heterostructure.
Normalized PL intensity maps of (c) MoS$_2$ and (d) MoSe$_2$.} \label{heterostruture}
\end{figure*}

We show a micrograph of the fabricated heterostructure in Fig.\ \ref{heterostruture}(a). It consists of a monolayer MoSe$_2$
flake, indicated by a blue dashed line, partially covered by a MoS$_2$ monolayer (white dashed line). The nominal angle between
the MoS$_2$ and MoSe$_2$ crystal axis is $\sim\SI{0}{\degree}$. Second harmonic spectroscopy confirmed that the actual stacking angle is $\sim\SI{1}{\degree}$, as discussed in the Supporting Information. Both layers are fully encapsulate in hBN and this van der Waals
stack is deposited on a SiO$_2$ substrate. A more detailed description of the fabrication procedure is provided in Methods.
Additional microscope images of the flakes used to prepare the heterostructure are shown in the Supporting Information. A
representative, broad range PL spectrum of our heterostructure at $T=5$\,K is presented in Fig.\,\ref{heterostruture}(b). The
peaks at \SI{1.615}{\eV} and 1.644\,eV arise from the radiative recombination of the intralayer trion and exciton of MoSe$_2$,
respectively. The weaker PL peaks centered at \SI{1.94}{\eV} and \SI{1.87}{\eV} are assigned to the recombination of neutral
excitons and to excitonic complexes bound to defects in MoS$_2$ \cite{surrente2017defect,cadiz2017excitonic}, respectively. The
low energy peak at 1.37\,eV stems from the interlayer exciton emission, formed due to the spatial separation of charge carriers in a type II heterostructure, as schematically shown in the inset of Fig.\ \ref{heterostruture}(b) \cite{baranowski2017probing, surrente2018intervalley}. In
this work, we focus on the properties of intralayer excitons. PL intensity maps of intralayer MoS$_2$ and MoSe$_2$ are
presented in Fig.\ \ref{heterostruture}(c,d), respectively. The MoSe$_2$ PL intensity of Fig.\ \ref{heterostruture}(d) is
suppressed in the heterostructure area, due to the interlayer charge transfer \cite{ceballos2014ultrafast, chen2016ultrafast}. The strongly modified optical properties in the heterostructure area are an indication of a good coupling between the layers, characteristic for \SI{0}{\degree} stacking angle \cite{wang2016interlayer}.

\begin{figure*}[!ht]
\centering
\includegraphics[width=1\linewidth]{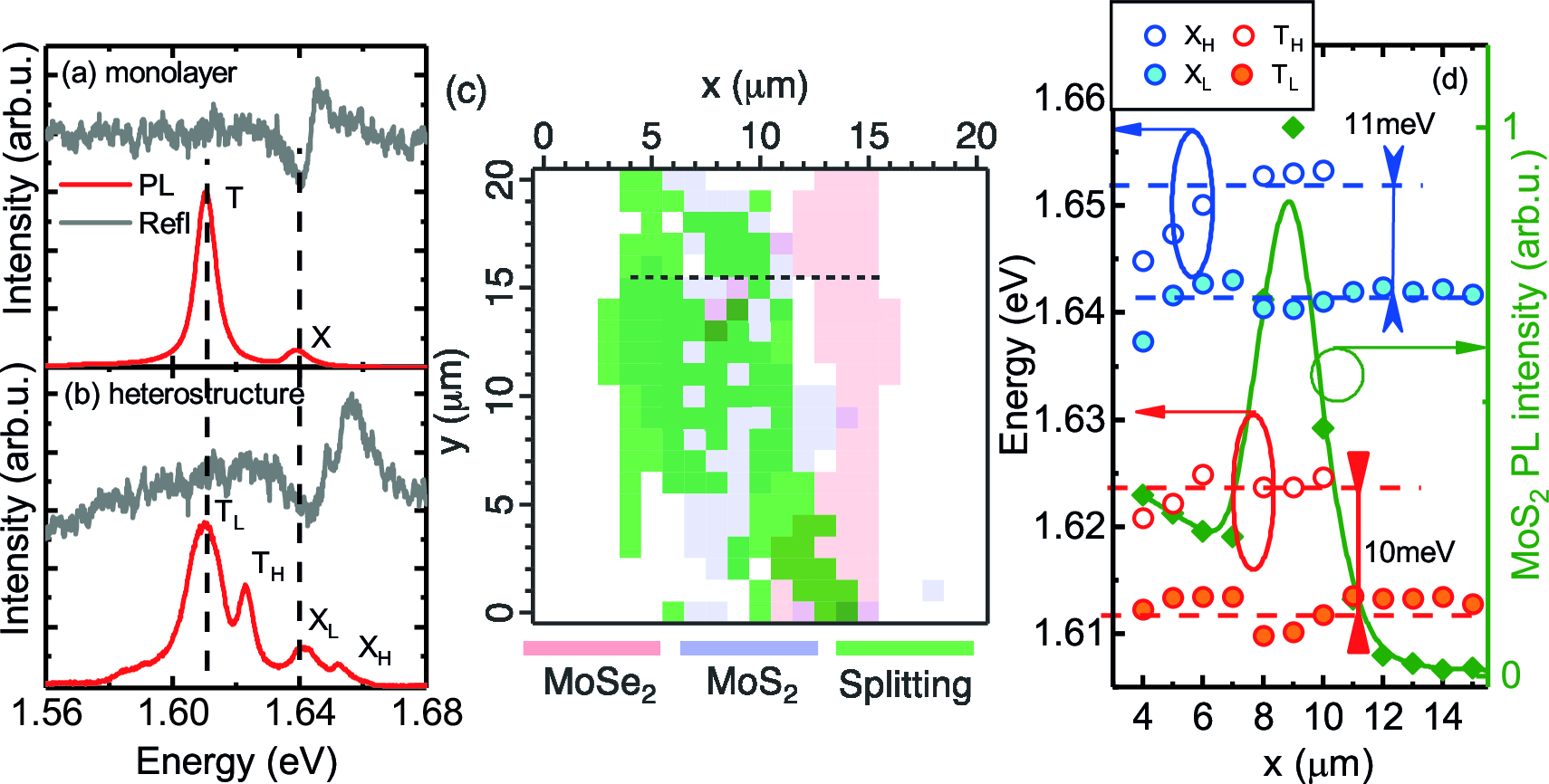}
\caption{PL and reflectivity spectra of MoSe$_2$  measured in (a) monolayer and (b) heterostructure. (c) Spatial map showing the
presence of MoSe$_2$ transition splitting, overlaid with the areas of most intense PL of MoS$_2$ and MoSe$_2$. (d) Energy of the
observed MoSe$_2$ transitions (blue and red dots) extracted along the horizontal dashed line of panel (c). Open circles
correspond to high energy peaks observed only in the heterostructure, closed circles represent low energy transitions observed in
all the MoSe$_2$ flake. Green diamonds:  MoS$_2$ PL intensity.} \label{doublet}
\end{figure*}

On a fully hBN-encaspulated MoSe$_2$ monolayer, the PL spectrum consists of two peaks, attributed to charged and neutral exciton (see Fig.\
\ref{doublet}(a)). In contrast, the MoSe$_2$ exciton and trion PL from the heterostructure region reveals a double peak
structure, as shown in Fig.\,\ref{doublet}(b). We label these features X$_{\text{L}}$ (T$_{\text{L}}$) and X$_{\text{H}}$
(T$_{\text{H}}$), to refer to higher and lower energy transition of the exciton (trion). A similar double structure is also
observed for the MoS$_2$ free exciton emission (see Supporting Information). Importantly, the splitting of the exciton is also
visible in the reflectivity spectrum of Fig.\ \ref{doublet}(b), unequivocally demonstrating that it is related to a free exciton
transition rather than to the emission from excitons bound to defects. The presence of a double PL peak in the heterostructure region is consistent with
the expected effect of the moir\'e pattern on the intralayer excitonic complexes. According to theoretical predictions, the
stacking of two lattice mismatched TMD monolayers induces a spatially periodic fluctuation of the potential, felt by the excitons,
\cite{yu2017moire,zhang2017interlayer,wu2018theory,Wu2017} with local minima related to different atomic registries
\cite{yu2017moire}. The period of the moir\'e pattern is in the range of a few to tens of nanometers (see Supporting Information for a discussion specific to our heterostructure). Therefore, the spatial
resolution of our far field optical measurements is not sufficient to resolve the spatial variation of the emission energy of
excitons located at different potential minima. Nevertheless, the emission from different optically active minima of the moir\'e pattern can be
spectrally resolved in our high quality sample. The double structure of the exciton and trion peaks is consistently observed only
in the heterostructure, while it is completely absent in monolayer regions. This is summarized in Fig.\ \ref{doublet}(c), where
the spatial map highlights the areas where the splitting is observed. To verify that the observed splitting is a direct consequence of the moir{\'e} pattern, we prepared two control samples (presented in Supporting Information), which consist of MoS$_2$/MoSe$_2$ heterostructures with stacking angles of $\sim\SI{20}{\degree}$ and $\sim\SI{60}{\degree}$. In former sample, the period of the moir{\'e} pattern is significantly smaller than the spatial extension of the intralayer exciton wave function and, as expected, we did not observe any splitting in the PL spectrum of the MoSe$_2$ exciton or trion. In latter sample, we systematically observe the splitting of the exciton and trion PL peaks of MoSe$_2$, whenever the excitation is performed within the heterostructure area. This is consistent with theoretical predictions \cite{yu2017moire}, and with the long range spatial period of the moir{\'e} pattern formed in the case of  $\sim\SI{0}{\degree}$ and $\sim\SI{60}{\degree}$ stacking angle (see Supporting Information).

In Fig.\ \ref{doublet}(d), we show the spatial variation of the
exciton and trion PL energy along the dashed line crossing the heterostructure in Fig.\ \ref{doublet}(c). The slice starts and
ends at positions on the sample where PL is observed. A direct comparison of the spectrum measured in the heterostructure region with one measured outside (see Fig.\
\ref{doublet}(a,b)) reveals that the double structure results from the appearance of a new energy peak on the high energy side of
the main exciton and trion emission. The energies of the emission lines observed in isolated monolayers are nearly identical to
the X$_{\text{L}}$ and T$_{\text{L}}$ lines observed in the heterostructure, as highlighted by the vertical lines in Fig.\
\ref{doublet}(a,b). This is in agreement with the prediction that the exciton states related to the moir\'e pattern should appear
as peaks on the high energy side of the main PL peaks \cite{yu2017moire,Wu2017}. This observation, along with the fact that the
double peak structure appears over a vast area in the heterostructure and not only at its edges \cite{raja2017coulomb}, allows us
to rule out the variation of the dielectric screening related to the presence of the MoS$_2$ monolayer as the origin of the
observed doublet structure. Since the dielectric constant of MoS$_2$ is higher than that of hBN \cite{laturia2018dielectric}, a
dielectric screening effect would lead to the appearance of lower energy peaks when the excitation is performed on the
heterostructure \cite{raja2017coulomb}. We also exclude a strain-related origin of the new peak by noting that the red shift of the A$_{1\text{g}}$ mode of MoSe$_2$ in the heterostructure would suggest tensile strain of MoSe$_2$ comprised in the heterostructure, which is not compatible with the appearance of a high energy peak in the PL spectrum \cite{conley2013bandgap,lloyd2016band, castellanos2013local}  (see Supporting Information for more details). Finally, the observed energy scale of the PL peaks splitting ($\sim 6$--\SI{12}{\milli\eV}) is in agreement with the energy scale of potential variations predicted for the related
MoS$_2$/WSe$_2$ heterostructure \cite{yu2017moire}.

\begin{figure}[!ht]
\centering
\includegraphics[width=1\linewidth]{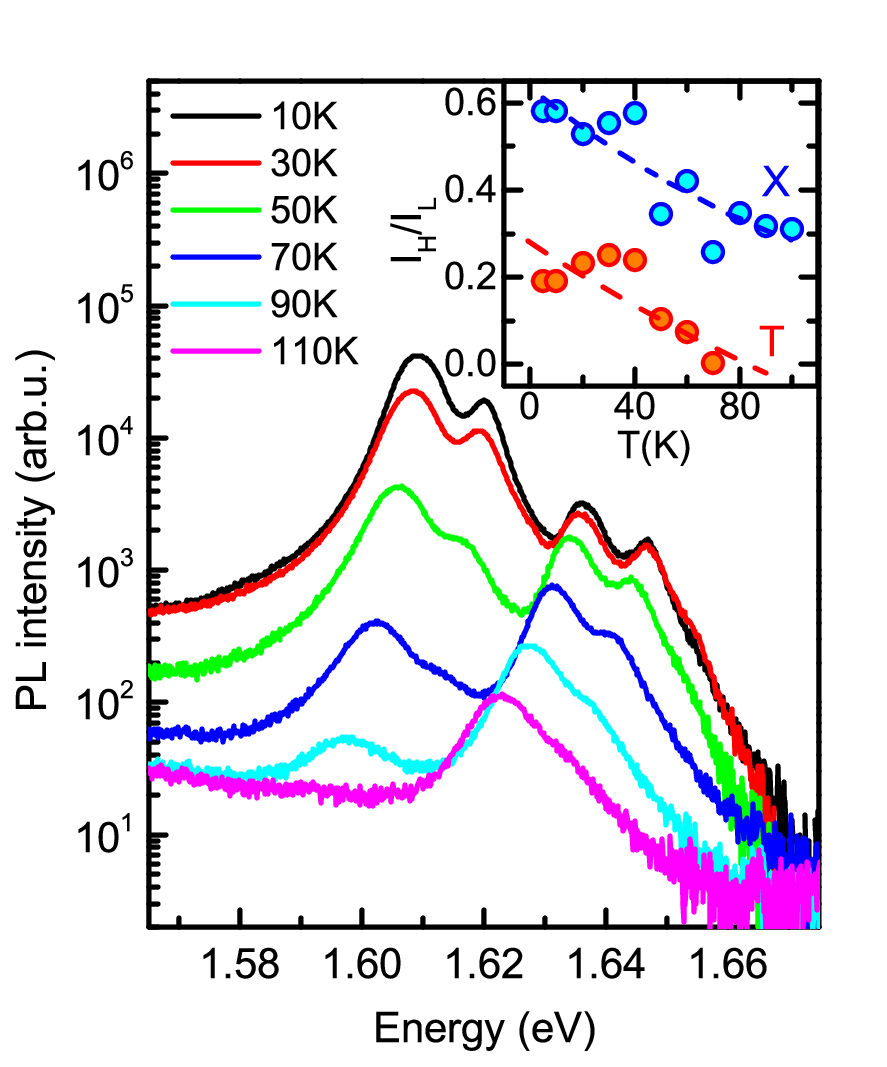}
\caption{Temperature dependence of  MoSe$_2$ PL. Inset: ratio of intensities of the high to low energy exciton and trion $I_{\text{H}}/I_{\text{L}}$.}
\label{temperature}
\end{figure}

In Fig.\,\ref{temperature}, we summarize the temperature dependence of the MoSe$_2$ PL measured on the heterostructure. As for
all Mo-based TMDs, we observe a decrease of the PL intensity with increasing temperature \cite{zhang2015experimental}. We notice in particular a faster decrease of the intensity of the trion PL, which can be no longer resolved at temperatures larger than $\sim\SI{100}{\K}$ \cite{surrente2017defect}, consistent with the smaller binding energy of this complex as compared to the exciton. Concerning the states split by the moir{\'e} potential, the intensity of the high energy peaks X$_{\text{H}}$ and T$_{\text{H}}$ decreases more rapidly than that of the
low energy peaks. For temperatures higher than $\sim\SI{90}{\K}$, corresponding to a thermal energy of $\sim\SI{7.7}{\milli\eV}$, the X$_{\text{H}}$ and T$_{\text{H}}$ features are no longer
resolved. This is illustrated in the inset of Fig.\,\ref{temperature}, where we plot the ratio of the intensities of the high
energy peaks $I_{\text{H}}$ normalized by the intensities of the corresponding low energy peaks $I_{\text{L}}$. The faster quenching of the X$_{\text{H}}$ and
T$_{\text{H}}$ emission further supports the moir\'e pattern as the origin of the trion and exciton doublet. A smaller
confinement is expected for the higher energy states \cite{yu2017moire}, hence with increasing thermal energy these excitons can be detrapped more easily via thermally activated phonon scattering out of the minima of the moir{\'e} potential. From the temperature dependence of the intensity of X$_{\text{H}}$, an activation energy of $\sim\SI{26}{\milli\eV}$ can be extracted (see Supporting Information). This value compares to the depth of the moir{\'e} potential for intralayer excitons of 12--\SI{19}{\milli\eV}, depending on the stacking of the heterobilayer \cite{yu2017moire}. This behaviour is fully consistent with the observation of a more rapid thermal quenching of the PL of high energy moir{\'e} states observed in the interlayer exciton \cite{tran2018moir}.

\begin{figure*}[!ht]
\centering
\includegraphics[width=1\linewidth]{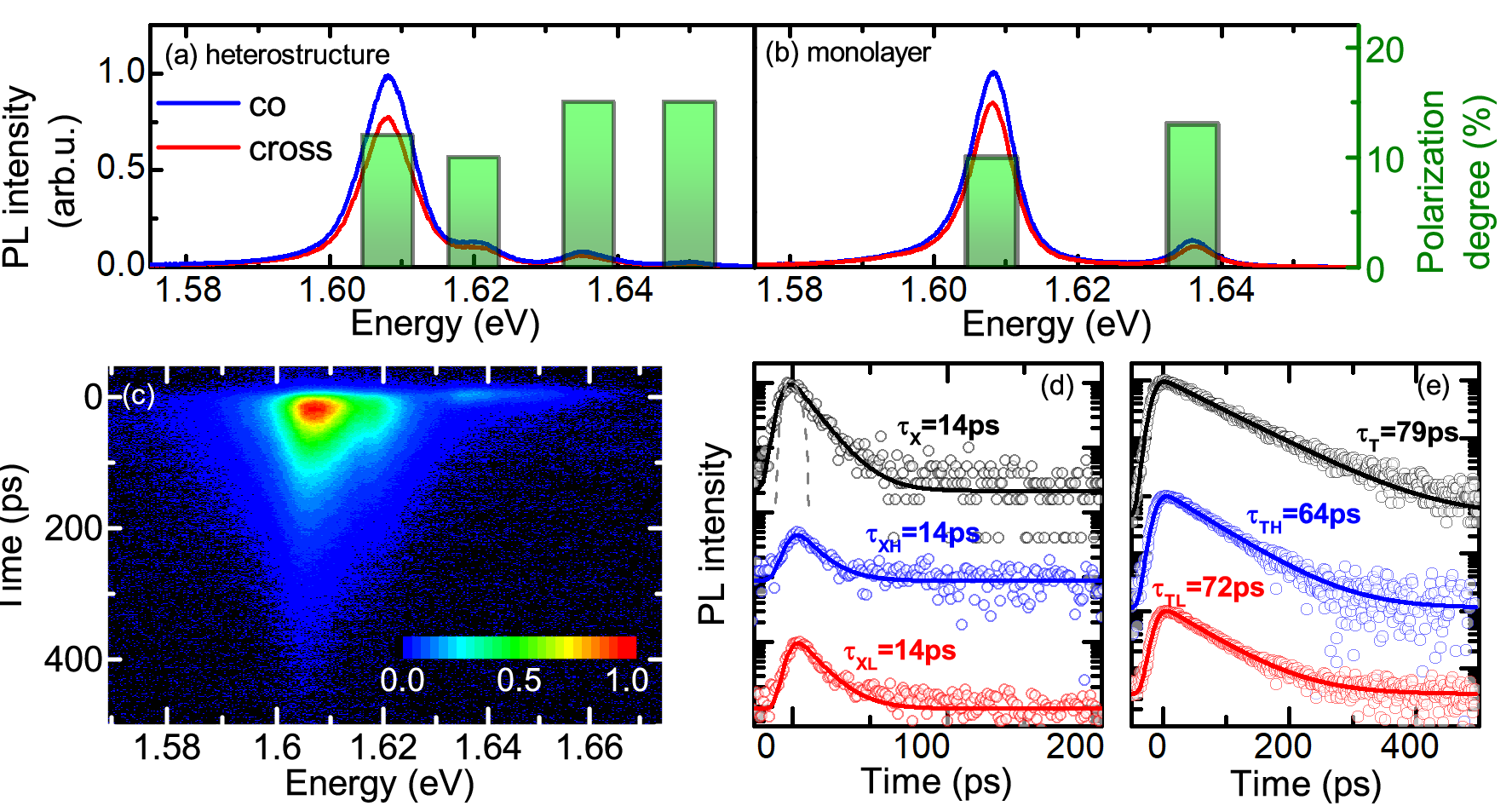}
\caption{Circularly polarization resolved PL spectrum of MoSe$_2$ measured (a) in and (b) out of the heterostructure. The green bars represent the degree of circular polarization of the peaks. (c) Spectrally resolved PL time evolution measured in the heterostructure area. (d) Decay curves (open circles) corresponding to low (red) and high (blue) energy of MoSe$_2$ (d) exciton and (e) trion from the heterostructure and decay of exciton PL outside heterostructure (black) together with fitted monoexponential decays. The extracted decay time constants are specified above the corresponding curves. The instrument response function is traced in dashed line in panel (d).} \label{polarization}
\end{figure*}

We investigated the valley polarization properties of the intralayer MoSe$_2$ excitons by exciting the PL with circularly
polarized light and detecting the co-polarized and cross-polarized circular polarization PL components. We show the PL spectra of
the isolated monolayer and of the monolayer comprised in a heterostructure in Fig.\ \ref{polarization}(a,b). The degree of
circular polarization $P_{\text{c}}$ of the PL, $P_{\text{c}}=(I_{\text{co}}-I_{\text{cross}})/(I_{\text{co}}+I_{\text{cross}})$,
is always positive (see green bars in Fig.\ \ref{polarization}). The degree of circular polarization is very similar for the
heterostructure and monolayer regions and amounts to about 10\% and 13\% for the trion and the exciton, respectively. There is no
significant difference between the polarization of high and low energy transitions in heterostructure. This suggests that,
consistent with theoretical predictions \cite{yu2017moire}, the selection rules for the intralayer exciton transitions are not
influenced by the presence of the moir\'e pattern, thus all transitions have the same polarization. This can be explained by
considering that the rotational symmetry of the transitions does not change in the plane of the monolayers, in contrast to the
interlayer transition for which the helicity of the emitted light varies across the moir\'e pattern \cite{yu2017moire,
wu2018theory}. The observed significant polarization is surprising considering that we excite far from excitonic transitions of
MoSe$_2$ \cite{wang2015polarization,kioseoglou2016optical,baranowski2017dark,surrente2017defect}. This effect might be related to
the encapsulation in hBN because we observe a similar degree of circular polarization in both the heterostructure and monolayer
regions, although the exact reason the relatively high degree of polarization after encapsulation requires additional
investigation.

Finally, we investigate the dynamics of the trion and exciton in MoSe$_2$. The spectrally resolved temporal evolution of the PL
measured in the heterostructure area is shown in Fig.\ \ref{polarization}(c). Similar data acquired on an isolated MoSe$_2$
monolayer is shown in the Supporting Information. In the streak image of Fig.\ \ref{polarization}(c), the four PL peaks are well
resolved, with the dynamics of the excitonic transition being significantly faster than those of the trions. To obtain more
quantitative information, we extract the decay curves corresponding to MoSe$_2$ trion and exciton transitions in monolayer and
heterostructure regions in Fig.\,\ref{polarization}(d,e), respectively. PL decays of all trion and exciton species can be fitted
well using a single exponential, convoluted with a Gaussian curve to account for the instrument response function of the system.
From this fitting, we extract the PL life times. There are no significant differences between PL decay times in the monolayer and
heterostructure areas or between X$_{\text{H}}$ (T$_{\text{H}}$) and X$_{\text{L}}$ (T$_{\text{L}}$) transitions. For the exciton
transitions, the PL decay time is $\sim 14$\,ps, while for the trion it is $\sim 70$\,ps. We notice, however, that the PL decay time of the exciton is close to the resolution of our system, therefore the dynamics of the two exciton states might be slightly different. The faster decay of the high energy trion is consistent with the observation on the dynamics of high energy interlayer exciton states in a moir{\'e} potential, where this behaviour has been attributed to the possible relaxation of high energy states to the low energy states \cite{tran2018moir}.

The overall similar decay times of the high and low energy transitions suggest that the moir\'e potential does not affect significantly the oscillator strength of the
transitions. This can be understood as a result of the slow spatial variation of moir\'e potential (a few to tens of nm, depending on the stacking angle and the lattice mismatch of the heterobilayer) compared to the exciton size in TMD monolayers ($\sim 1$\,nm) \cite{stier2018magnetooptics,stier2016probing}. Therefore, the wave function of
intralayer excitons is not significantly affected by the confinement induced by the moir\'e pattern. Interestingly, despite a clear drop of the PL intensity in the heterostructure area (Fig.\,\ref{heterostruture}), the decay times
of the trion and exciton PL are very similar to those observed in monolayer regions. This indicates that the charge transfer
between the layers occurs immediately after excitation and is effective only for hot carriers, in agreement with previous pump
probe measurements \cite{ceballos2014ultrafast}. The observed PL is related to thermalized excitons and the unchanged
decay times show that they are not affected by the interlayer transfer. This might be the result of the weak localization
potential induced by the moir\'e pattern. A larger intensity of the interlayer exciton PL has been observed in MoSe$_2$/MoS$_2$
heterostructures \cite{baranowski2017probing} with increasing temperature. This is probably the hallmark of the thermal
activation of excitons from shallow traps, represented by the moir\'e potential, followed by interlayer transfer. However,
further studies are needed to clarify this aspect, notably to distinguish it from a possible indirect character in k-space of the
interlayer transitions \cite{miller2017long, kunstmann2018momentum}.

In conclusion, we have demonstrated the impact of the moir{\'e} pattern formed in a MoS$_2$/MoSe$_2$ heterostructure on the
optical spectra of the intralayer exciton species. The potential fluctuations resulting from locally different atomic registries split the trion and exciton transitions of MoSe$_2$ and MoS$_2$ into two peaks related to optically active local minima of the
moir\'e potential. The nature of these doublets has been confirmed by detailed spatial mapping of the PL and by the temperature dependence of their PL. Polarization resolved measurements reveal that the selection rules of the transitions do not change with
respect to the case of isolated monolayers. The PL dynamics show that moir\'e induced potential does not change significantly
the oscillator strength of the transitions.  However, the moir\'e potential may be responsible for the suppression of the
interlayer transfer of thermalized excitons. All of the results presented are in agreement with the theoretically predicted
influence of moir\'e pattern on intralayer excitons in transition metal dichalcogenides.

\section*{Methods}
\textit{Sample preparation.} The samples have been prepared by all-dry deterministic transfer of mechanically exfoliated flakes
\cite{castellanos2014deterministic,frisenda2018recent}. hBN flakes have been exfoliated from commercially available hBN powder
(Momentive,  Polartherm grade  PT110) using Nitto tape (Nitto Denko corp. SPV 224). MoS$_2$ and MoSe$_2$ flakes have been also
exfoliated with Nitto tape from naturally occurring molybdenite mineral (Molly Hill mine, Quebec, Canada) and synthetic bulk
MoSe$_2$ grown by chemical vapour transport. The flakes cleaved from the bulk source of material are transferred onto a Gelfilm
(WF x4 6.0 mil by Gelpak) that is used as a viscoelastic stamp to perform the deterministic transfer. MoS$_2$ and MoSe$_2$
monolayers are selected from their transmittance and reflection spectra prior to their transfer \cite{frisenda2017micro}.
Monolayers with large faceted edges were used for the assembly of the heterostructures. The straight edges of the MoS$_2$ and
MoSe$_2$ were aligned with accuracy better than \SI{0.5}{\degree} during the transfer.

\textit{Optical spectroscopy.} For optical measurements, the sample was mounted on the cold finger of a helium flow cryostat with a quartz optical window.
Unless stated otherwise, all measurements have been performed at $T=5$\,K. Time resolved PL measurements were performed using a
CW frequency doubled solid state laser emitting at 532\,nm. The circularly polarization resolved PL was excited with the
frequency-doubled output of an optical parametric oscillator (OPO), synchronously pumped by a mode-locked Ti:sapphire laser and
tuned to \SI{600}{\nano\meter}. The temporal pulse width is typically \SI{300}{\femto\second}, with a repetition rate of
\SI{80}{\mega\hertz}. The instrument response function of the system, shown in Fig.\ \ref{polarization}(d), has a half width at half maximum of \SI{6}{\pico\s}. The excitation laser beam was focused on the sample using a $50\times$ microscope objective with a
numerical aperture of 0.55, giving a spot size of approximately \SI{1}{\micro\meter} diameter. The excitation power used for time-integrated measurements was \SI{20}{\micro\W}. The emitted PL was collected through the same objective and directed to a spectrometer
equipped with a liquid nitrogen cooled charge-coupled device (CCD) camera. For time-resolved measurements, the excitation was provided
by the OPO tuned to \SI{530}{\nano\meter}, using an average power of \SI{100}{\micro\W}. The collected signal was spectrally
dispersed using a monochromator and detected using a streak camera. Second harmonic measurements were performed in a similar setup. The Ti:sapphire laser was used as the excitation source. The laser light was polarized via a broadband polarizer and a halfwave plate was used to control the direction of the linear polarization of the fundamental wave. The laser was focused onto the sample using the same $50\times$ microscope objective used for PL measurements. The same set of halfwave plate and linear polarizer were used to analyze the polarization of the second harmonic signal, thus insuring that only the component of the second harmonic parallel to the polarization of the excitation was detected. The second harmonic signal was separated by the fundamental component by making use of a shortpass filter. The signal was finally detected by the spectrometer combined with the CCD.

\begin{suppinfo}
Micrograph of different stages of preparation of the van der Waals stack, second harmonic spectroscopy of the heterostructure presented in the main text, period of moir{\'e} pattern as a function of the stacking angle, \si{micro}PL spectra of additional devices with $\sim\SI{20}{\degree}$ and $\sim\SI{60}{\degree}$ stacking angle, activation energy of high energy exciton peak, Raman spectra of isolated MoSe$_2$ and MoSe$_2$ in a heterostructure, maps of the energy splitting of MoSe$_2$ exciton and trion, low temperature \si{\micro}PL spectrum of MoS$_2$ with split exciton peak, and time resolved PL of isolated MoSe$_2$.
\end{suppinfo}

\begin{acknowledgement}
This work was partially supported by BLAPHENE and STRABOT projects, which received funding from the IDEX Toulouse, Emergence
program,  ``Programme des Investissements d'Avenir'' under the program ANR-11-IDEX-0002-02, reference ANR-10-LABX-0037-NEXT. N.Z.
holds a fellowship from the China Scholarship Council (CSC). M.B. appreciates support from the Polish Ministry of Science and
Higher Education  within  the Mobilnosc  Plus program (grant no.\ 1648/MOB/V/2017/0). This project has also received funding from
the European Research Council (ERC) under the European Union's Horizon 2020 research and innovation programme (grant agreement
no.\ 755655, ERC-StG 2017 project 2D-TOPSENSE) and from the EU Graphene Flagship funding (Grant Graphene Core 2, 785219).
\end{acknowledgement}

\bibliography{Moire}

\begin{tocentry}

\includegraphics{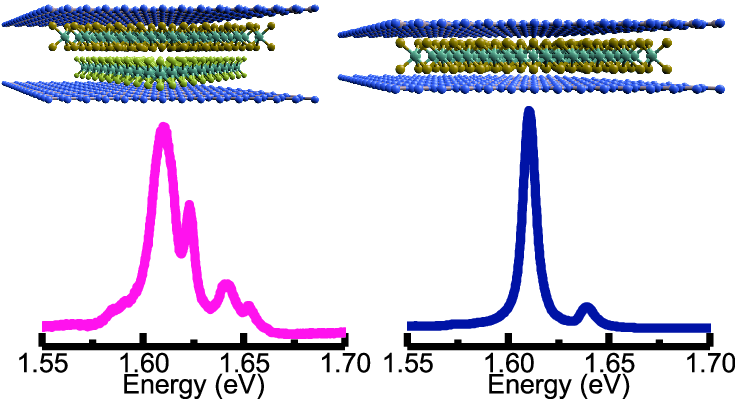}

\end{tocentry}

\end{document}